\documentclass[12pt]{iopart}
\input{psfig.sty}
\begin{document}

\title[Resonances in a trapped 3D  Bose-Einstein condensate]
{ Resonances in a trapped 3D  Bose-Einstein condensate
under periodically varying atomic scattering length}
\author{Fatkhulla Kh. Abdullaev\dag\ , Ravil M. Galimzyanov\dag
\footnote[7]{To
whom correspondence should be addressed (ravil@uzsci.net)}
, Marijana Brtka\ddag\ and Roberto A. Kraenkel\ddag\
}
\address{\dag\ Physical-Technical Institute of the Academy of
Sciences, 700084, Tashkent-84, G.Mavlyanov str.,2-b, Uzbekistan}

\address{\ddag\ Instituto de Fisica Teorica, UNESP, Rua Pumplona 145,
01405-900, Sao Paulo, Brasil}

\begin{abstract}
Nonlinear oscillations of a 3D radial symmetric Bose-Einstein condensate
under periodic variation in time of the atomic scattering length
have been studied analytically and numerically. The
time-dependent variational approach is used for the analysis of
the characteristics of nonlinear resonances in the oscillations of
the condensate.
The bistability in oscillations of the BEC width is invistigated.
The dependence of the BEC collapse threshold on the drive amplitude and
parameters of the condensate and trap is found.
Predictions of the theory
are confirmed by numerical simulations of the full Gross-Pitaevski equation.
\end{abstract}


\maketitle
\section{Introduction}
Investigation of collective excitations of a Bose-Einstein
condensate (BEC) has attracted a great attention for last years
\cite{Dalfovo}. Among the problems of these studies
a dynamical response of the condensate
to temporal perturbations in it as a whole is of natural interest.
Many efforts
have been devoted to the analysis of the effect of temporal modulations
in a trap potential on oscillations of the condensate. The parametric
and nonlinear resonances in BEC oscillations have been studied in
works \cite{Garcia,Garcia1,Stringari}. The dynamics of the BEC
under rapid strong
perturbations of the trap potential has been considered in
\cite{Dum,AbdGal}, where the stabilization of the condensate is shown
to exist. Since
the dynamics is nonlinear the chaotic oscillations are also
possible. Recently excitations of multimode dynamics and
stochastization in two component condensate oscillations were
observed experimentally in the work \cite{And}.
The collapse induced by the noise in the trap has been
studied in \cite{GAB}.

It should be noted that the effects of temporal modulations of a trap
potential on the condensate and noncondensate parts of the
cloud are similar.
Interesting nontrivial dynamics of the condensate is also observed
under temporal variation of the atomic scattering
length by means of the Feschbach resonances \cite{Kagan}. As
distinct from the trap modulations, such variations affect on the
condensed and thermal parts of an atomic cloud differently.
The dynamics of the condensate is
governed by the Gross-Pitaevskii equation with varying in time
nonlinear term. The temporal modulations of a trap potential leads
to the appearance of {\it time dependent linear potential} in the
nonlinear Schr\"odinger equation. As to the modulation of the scattering
length, $a_{s}$, it leads to the
Schr\"odinger equation with {\it time-dependent nonlinear
potential} that opens a new interesting area in the matter waves
theory. Here we can mention the Bosenova phenomenon occurring
under rapid change of the scattering length value from positive
(corresponding to the repulsion between atoms) to
negative (corresponding to the attraction between atoms)
\cite{Wieman}.

Periodic variation in time of the scattering length causes appearence
of new
effects. One of them is pattern formation in 2D  BEC  \cite{Stal}
(see related 1D problem in the context of the nonlinear optics
\cite{ADBS}). Another interesting phenomenon is the existence  of
resonances in the macroscopic quantum tunnelling between two
tunnel-coupled condensates \cite{AbdKra1}.
It should be yet mentioned a possible dynamical stabilization of 2D BEC under
rapid periodic variation
of the atomic scattering length \cite{Saito,ACKM} and existence of Feshbach
resonances.
Collective oscillations of a 1D Bose-Einstein gas under different external
perturbations of the trap and the scattering lenght have been considered
in the work \cite{Garnier}.

Oscillations of  2D BEC have been studied analytically and numerically
in the work \cite{AbdBronGal}. Numerical study of the 3D BEC oscillations
has been given in \cite{Adhikari}. At the same time analytical consideration
of oscillations of 3D BEC  is still lacking.
A free attractive 3D BEC is known to collapse at any number of atoms.
Existence of a trap leads to another situation. For a given trap potential
a 3D BEC goes to collapse if the number of atoms exceeds some critical value.
This has been checked experimentally in work \cite{Roberts}.

The purpose of this work is to investigate analytically and numerically
nonlinear oscillations of the 3D radial symmetric BEC under
periodic variations of the atomic scattering length. We will
employ the time-dependent variational approach and derive
a nonlinear ordinary differential equatin (ODE) for evolution of
the condensate width. Predictions for nonlinear resonances and
instability of
a condensate we will compare with numerical simulations of the full
time dependent Gross-Pitaevskii (GP) equation.

The article is organized as follows.
In section \ref{Sect:VarMod} we derive the variational model to describe
a 3D Bose-Einstein condensate in a trap potential. In section
\ref{Sect:PerVar} based on the Van-Der-Pole method
necessary equations are derived
to describe main and parametric resonances.
In section \ref{Sect:Results} numerical simulations as well as
their analysis
are given and finally in section \ref{Sect:Concl} the main
results are summarized.

\section{Variational Model}
\label{Sect:VarMod}
The total wave function for a 3D Bose-Einstein condensate in a trap
potential $V_{tr}(r)$ may be described by the Gross-Pitaevskii
equation:
\begin{equation}\label{NLSE}
i\hbar \Psi_t = -\frac{\hbar^2}{2m}\Delta \Psi + V_{tr}(r)\Psi +
g(t) |\Psi|^2 \Psi.
\end{equation}
Here $V_{tr}(r) = m\omega^2 r^2 /2$ is a trap potential, the factor
$g(t) = 4\pi \hbar^2 a_s/m$ and $a_s$ is the atomic scattering length.
We will suppose that the time-dependent scattering length $a_s$ is
constant to leading order, with
\begin{equation}
\label{as}
 a_s = a_0 \sigma(1 + h\sin(\Omega' t)),
\end{equation}
where $\sigma$ stands for the dimensionless scattering length.

Such variation can be obtained using for example varying in time
magnetic field. So the atomic scattering length can be expressed as
\begin{equation}
a_{s}(t) = a_{B} \left(1 + \frac{\Delta}{B_{0}-B_{1}(t)} \right),
\end{equation}
where $a_{B}$ is the asymptotic value of $a_{s}$ and the magnetic field
$B_{1} = B_{2}\sin{\Omega t}$. Then far from the resonance, we can approximate
$a_{s}(t) = a_{so}\sigma (1 + h\sin(\Omega t)),$
with $$a_{s0} = a_{B} \left( 1 + \frac{\Delta}{B_0} \right) ,
\ h = \frac{a_{B} \Delta B_{2}}{B_{0}^2 (1 + \frac{\Delta}{B_0})}.$$
As to experimental values of the parameters $B_0$ and $\Delta$
Feshbach resonances have been observed in $Na$ at 853 and 907 G
\cite{Inouye}, in $^{7}Li$ at 725 G \cite{Strecker} and in $^{85}Rb$
at 164 G with $\Delta = 11$ G \cite{Court}.
Another way of making variations in the nonlinear term is to vary the tight
confined potential
parameters of the low dimensional system. For a 2D pancake BEC
it may be the frequency of trap in the direction of tightening
and the transverse trap frequency in the case of a 1D cigar-shaped BEC.

Under the scaling
$\tau = t\omega$, \ $\rho = r /l_{0}, l_{0} = \sqrt{\frac{\hbar}{m\omega}}$,\
$U = \Psi \sqrt{4 \pi a_{s}l_{0}^{2}}$,
the Gross-Pitaevskii equation can be written in the dimensionless form
\begin{equation}\label{nlse_d}
iU_{\tau}  + \frac{1}{2}\Delta U -
\frac{\rho^2}{2}U + \sigma (1 + h\sin(\Omega \tau)) |U|^2 U = 0,
\end{equation}
where $\Omega = \Omega'/\omega$
and $\sigma =1, \ \sigma =-1 $ corresponds to attractive or repulsive
interactions between atoms in the condensate respectively.

For the radially symmetrical case the governing equation takes the form
\begin{equation}\label{nlse}
iU_{\tau}  + \frac{1}{2}U_{\rho \rho} + \frac{1}{\rho} U_{\rho} -
 \frac{r^2}{2}U + \sigma (1 + h\sin(\Omega \tau)) |U|^{2} U = 0.
\end{equation}

Corresponding Lagrangian density is determined by the following expression
\begin{eqnarray}
\label{lagr}
\fl L (U,U^*) = \frac{i}{2}(U^{*}U_{\tau} - U {U^*}_{\tau}) -
\frac{1}{2}|U_{\rho}|^2 - \frac{1}{2} \rho^2 |U|^2 +
\frac{\sigma}{2}(1 + h\sin(\Omega\tau)) |U|^4.
\end{eqnarray}
In deriving our variational model we proceed from the anzatz
that the modulus of the total wave function, $|U|$, evolves in a self-similar
way. So a trial function may be taken in the form
\begin{equation}\label{anz}
U(\rho,\tau) = A(\tau) R \left(\frac{\rho}{a(\tau)} \right)
\exp\left(i \frac{b(\tau) \rho^2}{2a(\tau)} + i \phi(\tau)\right) ,
\end{equation}

where the function $R^2 \left(\frac{\rho}{a(0)} \right)$ describes
initial BEC density distribution.

To obtain equations for the wave packet parameters
$A(\tau),a(\tau),b(\tau),\phi(\tau)$ we calculate the averaged
Lagrangian
$\bar{L}(\tau) = \int d^3 {\bf \rho}
L \left(U({\bf \rho},\tau),U^*({\bf \rho},\tau)\right)$

with the trial function (\ref{anz}). Its explicit
expression takes the form
\begin{eqnarray}
\label{lagr1}
\fl \bar{L} = -\frac{A^{2}a^{3}}{2}
\left( (b_{\tau}a - a_{\tau}b + a^2 + b^2)I_{2} +
2\phi_{\tau}n + \frac{I_4}{a^2} -
\sigma(1 + h\sin(\Omega\tau)) A^2 I_3 \right).
\end{eqnarray}
Hereafter the following designations for the constants
$n   = \int {R(\rho)}^2 \rho^2 d \rho $,
$I_2 = \int {R(\rho)}^2 \rho^4 d \rho $,
$I_3 = \int {R(\rho)}^4 \rho^2 d \rho $ and
$I_4 = \int {R_{\rho}(\rho)}^2 \rho^2 d \rho $
will be used.

The Euler-Lagrange equations for the functional $\bar L(\tau)$
lead to the following equations for the
scaling parameter $a(\tau)$ and chirp $b(\tau)$ :
\begin{eqnarray}
a_{\tau} = b , \nonumber\\
b_{\tau} = \frac{I_4}{I_{2}a^3} - a -
\frac{3\sigma (1 + h\sin(\Omega\tau)) A^2 a^3 I_3}{2a^4 I_2} .
\end{eqnarray}
Eliminating the parameter $b$ from the above system
and making substitution
$a = \left(\frac{I_4}{I_2}\right)^{\frac{1}{4}}v$
we get the following evolution equation for $v$ :
\begin{equation}\label{ode}
v_{\tau \tau} = \frac{1}{v^3} - v - \frac{\sigma P(t)}{v^4} ,
\end{equation}
where
\begin{eqnarray}\label{P}
P(\tau) = P_0 + P_1\sin(\Omega\tau) \, ,
P_{0}= \frac{3\sigma N_1 I_3 {I_2}^{\frac{1}{4}}}
{2n {I_4}^\frac{5}{4}} \, , P_{1} = P_{0} h
\end{eqnarray}
and the quantity
\begin{equation}\label{norm}
N_{1} = \frac{1}{4\pi}\int d^{3}{\bf r} |U|^2  =
A^{2}v^{3} \int r^2 dr R^2 = A^{2}v^{3} n = \frac{Na_{s}}{l_{0}},
\end{equation}
determines the norm of the wave function $U$.

It is easy to see that the above is a Hamiltonian system, with the
scaling parameter $v$ playing the role of a position variable,
and the chirp $b$ the conjugate momentum.
Then this equation can be also rewritten as \cite{Garcia1}
\begin{equation}\label{odepot}
v_{\tau \tau} = - \frac{\partial}{\partial v} V_{eff}(v) ,
\end{equation}
where
\begin{equation}\label{Veff}
V_{eff}(v) = \frac{1}{2v^2} + \frac{1}{2} v^2 -
\frac{\sigma P}{3 v^3} .
\end{equation}

The dynamics of a condensate is different for the cases of positive and
negative atomic scattering lengths.
We will analyze these cases separately. First, for sake of completeness,
let us consider $P_1 = 0$, that is, the case when the scattering length
is constant.

\subsection { Positive scattering length }
\label{Subsect:PosSc}
When $\sigma = -1$, i.e. $a_{s0} > 0$, the equilibrium points $v_{0}$ of Eq.(\ref{ode})
satisfy:
\begin{equation}
\label{sirina1d}
 v_{0}^{5}=v_{0}+P_{0}.
\end{equation}
This is a fifth-order algebraic polynomial problem that has only
one positive real root. Its solution corresponds to a stable
equilibrium point.
 When the interactions are strong $P_{0}\gg1$, one can neglect the
 term  $v_{0}$ and obtain $v_{0}^{(0)}=P_{0}^{1/5}$.  After substitution of
 $v_{0}=v_{0}^{(0)}+\delta$ into (\ref{sirina1d}), we obtain
\begin{equation}
\label{poz}
 v_{0}=\frac{5P_{0}}{5(P_{0})^{4/5}-1} .
\end{equation}
This corresponds
 to the Thomas-Fermi regime, since we neglect the quantum pressure in comparison with
parabolic potential and nonlinearity terms in the GP equation.

The frequency of small oscillations near this point is
\begin{equation}
\omega_{r} \approx \sqrt{5} \left(1 + \frac{3}{10 P_{0}^{4/5}} \right) ,
\end{equation}
where it is assumed that $P_{0} >> 1$.

\subsection { Negative scattering length }
\label{Subsect:NegSc}

A typical behavior of $V_{eff}(v)$ for two cases,  attractive
and repulsive condensates, are presented in figure \ref{fig_1}.


The equation for the equilibrium position
$$ v_{0}=\frac{1}{v_{0}^{3}}-\frac{
P_{0}}{v_{0}^{4}} , \eqno{(\ref{sirina1d}')}$$ has either no, or two
positive real solutions. The critical value of $P_{0}$ is given by
\begin{equation}
\label{Pkrticno}
 P_{c}=\frac{4}{5^{5/4}}=0.5350.
\end{equation}
When $P_{0} > P_{c}$, there are no equilibrium points, and the
condensate will collapse \cite{Saito1}. When $P_{0} < P_{c}$ there are two
equilibrium points, one of them unstable ($v_{01}$) and the other
one stable ($v_{02}$). This means that collapse can be avoided
when two conditions are satisfied:
\begin{itemize}
\item [a)] $P_{0} < P_{c} $
 \item [b)] the initial condition: $v_{03}>v>v_{01}$, where $v_{03}$
 is defined by: $V(v_{03})=V(v_{01})$.
\end{itemize} In this region the motion should be periodic motion around the
equilibrium point.

 When $ P_{0}  \ll1$ for the equilibrium point of (\ref{ode})
(the minimum of the potential $V_{eff}$, see figure \ref{fig_1})
we have the following approximation
 \begin{equation}
 \label{neg}
 v_{0}=1-\frac{P_{0}}{4} .
\end{equation}
In the interval $0 < P_{0} < 0.5$
the position of maximum of the potential function $V_{eff}$ is
well approximated by $v_{01} \approx P_{0}$, the point $v_{03} \approx
1/(\sqrt{3}P_{0})$. The frequency of small oscillations near
$v_{0}$ is
\begin{equation}
\omega_{a} \approx 2\sqrt{1 - \frac{P_{0}}{4}} .
\end{equation}

\section { Periodic variation in scattering length }
\label{Sect:PerVar}
One of the problems of interest in BEC is the study of the
oscillation in the condensate subject to a periodic variation of
the scattering length \cite{AbdKra1,AbdBronGal,Adhikari,Stal}.
Such a variation, as was above mentioned, can be achieved
experimentally by varying the magnetic field or using optically
induced Feshbach resonances. In GP equation the temporal
modulation of the scattering length of atoms corresponds to a
time-dependent nonlinear coefficient.

For the sinusoidal variations of the atomic scattering
length (\ref{P}) the
equation (\ref{ode}) takes the form:
 \begin{equation}
\label{1dp1}
 \frac{d^{2}v}{d\tau^{2}}+v=\frac{1}{v^{3}}-
\frac{\sigma P_{0}}{v^{4}}-\frac{\sigma P_{1}}{v^{4}}\sin(\Omega t).
\end{equation}
 We are looking for a solution of this equation in the form
$$v=v_{0}+v_{1},$$
where $v_{1}\ll v_{0}$ is a small deviation from the equilibrium
position $v_{0}$.
Substituting it into (\ref{1dp1}) and expanding the terms of the equation
near the equilibrium point $v_{0}$  we get the equation for $v_{1}$:
\begin{equation}
\label{velika}
\ddot{v}_{1}+\omega_{0}^{2}v_{1}+Av_{1}^{2}+Bv_{1}^{3}+... =
(C+Dv_{1}+Ev_{1}^{2}+Fv_{1}^{3}+...)\sin(\Omega t).
\end{equation}
Above, we introduced the following constants:
\begin{eqnarray}
\fl \omega_{0}^{2}=1+3/v_{0}^{4}-4\frac{\sigma P_{0}}{v_{0}^{5}},
A=-6/v_{0}^{5}+10 \sigma P_{0}/v_{0}^{6},\
B=10/v_{0}^{6}- 20 \sigma P_{0}/v_{0}^{7}, \nonumber \\
\fl C=-\sigma P_{1}/v_{0}^{4},\ D=4\sigma P_{1}/v_{0}^{5},\
E=-10 \sigma P_{1}/v_{0}^{6},\
F=20 \sigma P_{1}/v_{0}^{7}. \nonumber
\end{eqnarray}
In addition, we will assume that $v_{1}$ and $P_{1}$ are  small
quantities of the same order, say $\varepsilon$. (Note that
constants $C$, $D$, $E$ and $F$ depend on $P_{1}$.) Then, we can
keep in equation (\ref{velika}) terms up to order $\varepsilon^{3}$:
\begin{equation}
\label{glavna}
\ddot{v}_{1}+\omega_{0}^{2}v_{1}+Av_{1}^{2}+Bv_{1}^{3}=
(C+Dv_{1}+Ev_{1}^{2})\sin(\Omega t).
\end{equation}

Now, we proceed on the examination of the resonance cases of
equation (\ref{glavna}). We assume that
$$\omega_{0}\approx\frac{p}{q}\Omega,$$ where $p$ and $q$ are
mutually prime integers.
Solution of equation (\ref{glavna}) may be sought in the form of the expansion
\begin{equation}
\label{v1}
 v_{1}=a\cos(\Omega t+\theta) + h x^{(1)} + h^2 x^{(2)} + ...  ,
\end{equation}
where $h$ is the amplitude of periodical perturbation (see equation (\ref{as})).
Making use of the successive approximations method described in
reference \cite{bogoljubov}, in the case of {\it main resonance} ($p=q=1$)
we get the following set of equations for the amplitude, $a$ and
phase $\theta$:
\begin{equation}
\label{main} \frac{da}{dt}= \left(-\frac{C}{2\Omega} -
\frac{E}{8\Omega}a^{2} \right)\cos(\theta),
\end{equation}

$$\frac{d\theta}{dt}=\omega_{0} - \Omega + \frac{C}{2a\Omega}\sin(\theta) +
\left( \frac{3B}{8\Omega} - \frac{5 A^2}{12 \omega_{0}^2 \Omega} \right) a^{2}
+ \frac{3E}{8\Omega}a\sin(\theta) .
\eqno{(\ref{main}'),}$$

For the steady-state regime
$da/dt=d\theta/dt=0$. Then the amplitude frequency response
can be easily found:
\begin{equation}
\label{AmpFreq}
\beta a^{3}\pm 3Ea^{2}\pm 4C+8\Omega\gamma a=0,
\end{equation}
where $\gamma=\omega_{0}-\Omega$, $\mid \gamma\mid \ll 1$ and
$$\beta = 3B - \frac{10A^2}{3\omega_0 ^2} .
\eqno{(\ref{AmpFreq}')}$$
The signs ($\pm$) in (\ref{AmpFreq}) corresponds to different
branches of the response (i.g. see figure \ref{AmFre}).

  System (\ref{main}) can be transformed into hamiltonian form by
changing the variable $a=\sqrt{Q}$. In this case $ dQ/dt=\partial
H/\partial \theta$,  and $d\theta/dt=-\partial H/\partial Q $,
where
\begin{equation}
\label{hamiltonijan}
H(Q,\theta)=-\frac{C}{\Omega}\sqrt{Q}\sin(\theta) -
\frac{E}{4\Omega}Q^{3/2}\sin(\theta)+
(\Omega-\omega_{0})Q-\frac{\beta}{16\Omega}Q^{2}.
\end{equation}

 In the case of {\it parametric resonance}
($p=1, q=2,v_{1}=a\cos(\frac{1}{2}\Omega t+\theta)$)
the equations for $a$ and $\theta$ have the form:
\begin{equation}
\label{parametrica}
 \frac{da}{dt}=-\frac{D}{2\Omega}a\cos(2\theta)
\end{equation}
\begin{equation}
\label{parametricteta}
\frac{d\theta}{dt}=\omega_0 - \frac{1}{2}\Omega +
\frac{D}{2\Omega}\sin(2\theta) +
\left(\frac{3B}{4\Omega} - \frac{5 A^2}{6 \omega_{0}^2 \Omega} \right)a^{2}
\end{equation}
Corresponding Hamiltonian is:
\begin{equation}
H_{p}(Q,\theta)=-\frac{D}{2\Omega}Q\sin(2\theta) +
\left(\frac{1}{2}\Omega-\omega_{0} \right)Q-
\frac{\beta}{8\Omega}Q^{2}.
\end{equation}

 It follows that $H$ is the integral of motion and there
 exists a separatrix $H=0$ separating finite trajectories
 corresponding to nonlinear resonances from infinite ones.

Phase planes for two cases of attractive BEC ($\sigma =1$)
with $P_0 = 0.2$ and $P_0 = 0.4$ are
shown in figure \ref{hamiltonian}
As seen the portraits differ drastically. The reason is different signs
of factor $\beta$ entering equation (\ref{main}) for the cases $P_0 = 0.2$ and
$P_0 = 0.4$.

Phase planes for two cases of repulsive BEC ($\sigma =-1$)
with $P_0 = 0.4$ and $P_0 = 0.8$ are
shown in figure \ref{ham_repulse}

\subsection{Dynamics of an attractive condensate under periodic $a_{s}$}

  Oscillations with the maximum amplitude for the case of
the main resonance
take place for separatrix $H_{s} \approx 0$ when   $\theta=\pi/2$
(see equation (\ref{hamiltonijan})).
The critical drive amplitude $P_{1c}$ providing the maximum amplitude of
the attractive condensate width oscillations, $a_{max}$
may be obtained from (\ref{AmpFreq}). In order to obtain an expression
for $P_{1c}$ let's rewrite above defined parameters $C$, $D$ and $E$
introducing new ones $c, d, e$ as
$C = P_{1} c$, $D = P_{1} d$ and $E = P_{1} e$. Then in the case of
$\Omega = \omega_0$ ($\gamma = 0$) we get
\begin{equation}
\label{P1critical}
P_{1c} = \frac{\mid \beta \mid a_{max}^2}{\mid 3e a_{max} +
\frac{4c}{a_{max}} \mid}.
\end{equation}
The maximum amplitude $a_{max}$ is found as follows.
As known the particle leaves the well (\ref{Veff}) when
its kinetic energy $\langle v_{1t}^{2}\rangle/2$ becomes
comparable with the difference between the potential well bottom
and separatrix, $\Delta V$. This energy is:
\begin{equation}
\Delta V = V_{eff}(v_{01}) - V_{eff}(v_0),
\end{equation}
where $v_0$ is stable and $v_{01}$ is unstable equilibrium points
(see subsection \ref{Subsect:NegSc}).
For the case of slowly varying amplitude $a$ and phase $\theta$,
we have from equation (\ref{v1}):
\begin{equation}
v_{1t}=-a \Omega\ sin(\Omega t + \theta).
\end{equation}
Equating critical value of the kinetical energy corresponding to the maximum
amplitude of oscillations $a_{max}$ to $\Delta V$ we arrive to the following
expression
\begin{equation}
\label{amax1}
a_{max} = \frac{\sqrt{2 \Delta V}}{\Omega}.
\end{equation}

Another way of calculating the maximum amplitude, $a_{max}$ is to employ
a direct expressiun for it
\begin{equation}
a_{max} = \frac{1}{2} (v_{03} - v_{01}),
\end{equation}
where the points $v_{03}$ and $v_{01}$ have been defined in

subsection \ref{Subsect:NegSc}.

In the region $0 < P_0 < 0.4$ where the well is deep,
we have the approximations $v_{01} \approx P_0$ and
$v_{03} \approx \frac{1}{\sqrt{3}P_0} - \frac{9}{2\sqrt{3}} P_0^3$.
Then in the indicated region of $P_0$ for $a_{max}$
we get a very good approximation
\begin{equation}
\label{amax2}
a_{max} \approx \frac{1}{2} \left(\frac{1}{ \sqrt{3}P_0} -
\frac{9}{2\sqrt{3}}P_0^3 - P_0 \right) .
\end{equation}

\section{ The results of numerical simulations of the full
Gross-Pitaevskii equation}
\label{Sect:Results}

Based on the above-stated theory we can now proceed to numerical
simulations to test calculations of the previous sections and
carry out more detailed analysis.
We discretize the problem
in a standard way, with time step $\Delta \tau$ and spatial step $\Delta r$,
so $U_j^k$ approximates $U(j\Delta r,k\Delta \tau)$. More specifically
we approximate the governing equation (\ref{nlse})
with the following second order accurate
semi-implicit Crank-Nicholson scheme,
\begin{eqnarray}\label{dnlse}
\fl \frac{i(U^{k+1}_{j} - U^{k}_{j})}{\Delta \tau} =
-\frac{1}{2\Delta r^2}
\left[(U^{k+1}_{j-1} - 2U^{k+1}_{j} + U^{k+1}_{j+1}) +
(U^{k}_{j-1} - 2U^{k}_{j} + U^{k}_{j+1})\right] - \nonumber \\
\frac{1}{2 r_j \Delta r}
\left[(U^{k+1}_{j+1} - U^{k+1}_{j-1}) + (U^{k}_{j+1} - U^{k}_{j-1})\right] +
\nonumber \\
\frac{1}{2}\left[\frac{{r_j}^2}{2}
(1 + h\sin(\Omega \tau_k)) \sigma |U^{k}_{j}|^2\right]
(U^{k}_{j} + U^{k+1}_{j}),
\end{eqnarray}
The set of algebraic equations (\ref{dnlse}) is solved by the
vectorial sweep method.

In solving the full governing partial differential equation (PDE)
(\ref{nlse})
both for  attractive BEC ($\sigma =+1$) and repulsive one ($\sigma =-1$)
we employ three different anzatzes as initial wave-packets :
anzatz 1, called a Gaussian one \cite{Anderson,Stoof,Garcia}, whith
the trial function in (\ref{anz}) choosen as
$$R\left(\frac{\rho}{a}\right) =
\exp\left(-\frac{1}{2}\left( \frac{\rho}{a}\right) ^2\right)$$
and anzatzes 2 and 3 when
the trial function $R(\frac{\rho}{a}) = R(\xi )$ where $R(\xi )$
is the solution of the equation
\begin{equation}\label{anzEq}
\frac{1}{2}\frac{d^2R}{d {\xi}^2} + \frac{1}{\xi}\frac{dR}{d \xi} -
\frac{{\xi}^2}{2} R + R^3 - \lambda R = 0
\end{equation}
with $\lambda$ equal to $-1$ and $+1$ correspondingly. Such a choice
enables us to use different initial wave packets.

The parameters of the initial wave-packet are chosen to provide
the given value of the norm $N$ and $P_0$
determined by Eqs.(\ref{norm}) and (\ref{P}).

In PDE simulations the current values of the scaling parameter $a(\tau)$ is
calculated as follows :
\begin{equation}
a(\tau) = a_0 \sqrt{ \frac{\int d\rho {\rho}^4 |U(\rho, \tau)|^2}
{\int d\rho {\rho}^4 |U(\rho, 0|^2}}. \nonumber
\end{equation}

Figure \ref{profiles} depicts profile of initial wave packets
$|U(\rho, \tau = 0)| $ corresponding to anzatzes 1, 2 and 3. Parameters
of the wave packets are chosen to give the same value of the system
parameter $P_0 = 0.2$.

%

Main wave packet parameters are given in Table below.
\vskip0.2in
\begin{tabular}{|l|c|c|}
\hline
$ wave packet $ & $ norm N $ & $ <r> $ \\\hline
 anzatz 1      &  0.01973    &  1.06250  \\
 anzatz 2      &  0.02254    &  1.07056  \\
 anzatz 3      &  0.01836    &  1.06698  \\\hline
\end{tabular}
\vskip0.2in

As will be seen further in spite of some difference in the wave packets
parameters their dynamics are atmost identical at small values of $P_0$.
It should be also noted that anzatz 1 gives exact solution at $P_0 =0$,
anzatz 2 does it at $P_0 = 0.372$ and anzatz 3 corresponds to a collapsing
solution.

\subsection { Attractive condensate ($\sigma = +1$)}

The results of numerical simulations of PDE and ODE models in the case
of {\it main resonance} are presented below.

Figure \ref{m_res} depicts dynamics of the scaling parameter $a$
of an attractive BEC under main resonance obtained from PDE and ODE
calculations for two cases with: a)  $P_0 = 0.2$, $P_1 = 0.014$ and
b) $P_0 = 0.35$, $P_1 = 0.00525$.
The frequency $\Omega $ of the periodical perturbation is taken to be equal
to the eigen frequency of the system $\omega_0$ determined by
\begin{equation}
\label{w0attr}
\omega_{a} = 2\sqrt{1 - \frac{P_{0}}{4v_{0}^5}} .
\end{equation}
In both cases of PDE simulations initial wave packet has gaussian form.
The agreement between simulations of the
full PDE and ODE in case a) is good enough for the time less than 100.
In case b) when  $P_0 = 0.35$ there is clearly some discrepancy
in the actual value of oscillation amplitude.
It may be
explained by the shift of the equilibrium point
in full PDE calculations.
%

In the case of {\it main resonance} the amplitude frequency response
of the scaling parameter resonant oscillations is described
by equation (\ref{AmpFreq}). The theoretical amplitude frequency response
based on this equation and the one obtained from full PDE simulations
for the case of $P_0 = 0.3$ and $P_1 = 0.0045$ are given in figure \ref{AmFre}.
In PDE calculations self similar anzatz (anzatz 2)
from equation (\ref{AmpFreq}) is used
as initial wave-packets. Each point of the plot results from PDE simulation
of the scaling parameter evolution under main resonance starting from the
stationary point $a_0$ where the oscillations amplitude is equal to zero.


The factor $\beta$ in  (\ref{AmpFreq}) plays an
important role in the amplitude
frequency response. Its sign determines the incline direction
in the plot of the response and its value determines degree of the
nonlinearity in the main resonance. Small values of the factor $\beta$
means prevalence of the linear regime in the resonance.

As was above shown collapse of the attractive BEC condensate
can be avoided when $P_0 < P_c$ (see (\ref{Pkrticno})). We have carried
out full PDE simulations
to study conditions when the main resonance makes the BEC to leave
the potential well $V_{eff}$ and the BEC begins to collapse.
In terms of the effective Hamiltonian (\ref{hamiltonijan}) this means
that when in the resonance the amplitude of oscillations exceeds the value
on the separatrix ($H > 0$), the condensate collapses.

The dependence of the critical drive amplitude $P_{1c}$ on $P_0$
obtained from ODE (solid line) and
PDE (filled squares)
numerical simulations as well as the theoretical dependence
(\ref{P1critical}) (dotted line) are shown in figure \ref{escape}.
The theoretical curve is obtained from equation (\ref{P1critical}) using
approximation (\ref{amax2}) at $0 < P_0 < 0.4$ and
expression (\ref{amax1}) at $0.4 \ge P_0 < P_{1c}$.
As seen ODE simulations based on the equation (\ref{1dp1})
is in a good agreement with PDE simulations of the full Gross-Pitaevski
equation (\ref{nlse}) where self similar anzatz 2 ($\lambda = -1$)
is used as an initial wave packet.

The dependences obtained from ODE and PDE simulations has a brightly
expressed minimum at the point $P_0 = 0.2$. It may be explained by
supposition that the {\it resonance} at this point is mainly {\bf linear}.
Indeed in accordance with equation (\ref{AmpFreq}') the factor $\beta$ entering
the folmula for the amplitude-frequency response goes to zero
at $P_0 \approx 0.277$. It means that at this point the resonance
is close to linear and small value of external drive $P_1$ leads to
large amplitudes of the BEC oscillations.
Discrepancy between positions of the minimum in theoretical curve and
the one obtained from numerical simulations may be explained
by the fact that in deriving the governing equation (\ref{velika})
the terms of higher degrees than 3 are not taken in account.
%

Next plot in figure \ref{pres1} depicts the dynamics of the {\it parametric}
resonance $(\Omega = 2 \omega _0)$ at the point $P_0 = 0.36$
when $P_1 = 0.051$. Full PDE calculations were carried out with initial
wave packet corresponding to anzatz 3 ($\lambda = 1$).
%

As seen at times  $t < 50$ parametric resonance is proved to be
suppressed in accordance with the prediction of the theory.

\subsection { Repulsive condensate ($\sigma = -1$)}

The dynamics of the scaling parameter $a$
of a repulsive condensate under main resonance obtained from PDE and ODE
calculations for the case $P_0 = 0.4$, $P_1 = 0.06$ is shown
in figure \ref{m_resrep}.
The frequency of the periodical perturbation $\Omega = \omega_0$,
where the eigen frequency $\omega_0$  is determined as
\begin{equation}
\label{w0rep}
\omega_{a} = 2\sqrt{1 + \frac{P_{0}}{4v_{0}^5}} .
\end{equation}
(for comparison see equation (\ref{w0attr})).
An initial wave packet has gaussian form (anzatz 1).
The agreement between simulations of the
full PDE and ODE  is excelent for the time less than 100.

The dynamics of the repulsive  BEC under {\it parametric resonance}
resonance $(\Omega = 2 \omega _0)$ obtained from full PDE calculations
for $P_0 = 0.4$ and
$P_1 = 0.12$ are shown in figure \ref{pres2}. Two cases are presented,
the case when the initial wave packet :  a) has self similar form
(\ref{anzEq}) with $\lambda = 1$ (anzatz 3)
and b) is gaussian (anzatz 3).
%

\section{Conclusion}
\label{Sect:Concl}

In this work we study the nonlinear oscillations of the 3D
spherical symmetric Bose-Einstein condensate under periodic
variation in time of the atomic scattering length. We employ the
time-dependent variational approach and derive the equation for
the dynamics of the condensate width. The equation has the form of
the unharmonic oscillator under the singular periodic drive. Using
this equation we analyze analytically the resonances
characteristics for the cases of attractive and repulsive
condensates . In particular we investigate the process of the attractive
condensate collapse  with $N< N_{c}$  under
periodic variations of the scattering length. We find the possibility
for the bistability in the oscillations condensate width  depending
on the detuning between drive frequency and the frequency of small
oscillations of a condensate.

We have shown that even when the number of the condensate atoms less
than the critical, the condensate can collapse. Using the variational
approach equation we derive the dependence of the driving force critical
value $P_{1c}$ versus initial parameter $P_0$.

For the repulsive condensate case we study both main and parametric
resonances in condensate oscillations. The full numerical simulations
of the Gross-Pitaevskii equation with periodic nonlinear coefficient
confirms predictions based on the variational approach.
We find the deviations from  predictions of the variational approach
for the case when  the number of atoms close to the critical one.

\ack
We acknowledge partial support by the FAPESP, Brasil.


\Figures

\begin{figure}
\caption{\label{fig_1}
Effective potential used in describing the scaling parameter $v$.
Two cases are presented : a) $P_0 = 0.4$ (attractive BEC) and
b) $P_0 = -0.4$ (repulsive BEC).
}
\end{figure}

\begin{figure}
\caption{\label{hamiltonian}
The phase planes for system (\ref{main}), (\ref{main}') at
(a) $P_0 = 0.2$ ($P_1 = 0.008$, $\Omega = 1.936$)
and (b) $P_0 = 0.4$ ($P_1 = 0.016$, $\Omega = 1.766$) for
$\sigma=1$ (attractive BEC),  $\gamma=0$ and different
$H$.}
\end{figure}

\begin{figure}
\caption{\label{ham_repulse}
The phase planes for the system (\ref{main}), (\ref{main}') at
(a) $P_0 = 0.4$ ($P_1 = 0.06$, $\Omega = 2.066$)
and (b) $P_0 = 0.8$ ($P_1 = 0.2$, $\Omega = 2.100$) for
$\sigma=-1$, (repulsive BEC) $\gamma=0$ and different
$H$.}
\end{figure}

\begin{figure}
\caption{\label{profiles}
Profile of initial wave packets
$|U(\rho, \tau = 0)|$ corresponding to anzatzes 1, 2 and 3. Parameters
of the wave packets are chosen to give the same value of the system
parameter $P_0 = 0.2$.
}
\end{figure}

\begin{figure}
\caption{\label{m_res}
Dynamics of the scaling parameter $a$ of an attractive BEC under main
resonance ($\Omega = \omega_0$) obtained from PDE and ODE calculations,
a) $P_0 = 0.2$, $P_1 = 0.014$ and b) $P_0 = 0.35$, $P_1 = 0.00525$.
Solid line is for PDE and dotted line for ODE calculations.
In both PDE simulations initial wavepackets are gaussian.
}
\end{figure}

\begin{figure}
\caption{\label{AmFre}
Amplitude frequency responses obtained from full PDE simulations
(dotted line) and from theory (solid line)
for the case when $P_0 = 0.372$, $P_1 = 0.00372$ and $\sigma = 1$
(attractive BEC).
In PDE calculations anzatz 2 from equation (\ref{AmpFreq}) is used
as initial wave-packets.
}
\end{figure}

\begin{figure}
\caption{\label{escape}
Dependence of the critical drive $P_{1c}$ leading to collapse on $P_0$.
Solid line is for ODE calculations and filled squares are for
full PDE calculations. Self similar anzatz (anzatz 2) with $\lambda = -1$
is chosen for initial wave packets in PDE simulations. Theoretical dependence
(\ref{P1critical}) is shown by dotted line.
}
\end{figure}

\begin{figure}
\caption{\label{pres1}
Dynamics of the scaling parameter $a$ under  {\it parametric resonance}
$(\Omega = 2 \omega _0)$ obtained from full PDE calculations
for $P_0 = 0.36$ and $P_1 = 0.051$.
As initial wave packet  self similar anzatz 3 ($\lambda = 1$) is used.
}
\end{figure}

\begin{figure}
\caption{\label{m_resrep}
Dynamics of the scaling parameter $a$ of a repulsive BEC under main
resonance ($\Omega = \omega_0$) obtained from PDE and ODE calculations,
$P_0 = 0.4$, $P_1 = 0.06$  ($\omega_0 = 2.0664$).
Solid line is for PDE and dotted line for ODE calculations.
An initial wavepacket is chosen to be gaussian.
}
\end{figure}

\begin{figure}
\caption{\label{pres2}
Dynamics of the repulsive BEC  under  {\it parametric resonance}
$(\Omega = 2 \omega _0)$
obtained from full PDE calculations
when $P_0 = 0.4$ and $P_1 = 0.12$.
Two cases are presented : a ) when self similar anzatz 3 is used as
an initial wave packet and b) when the initial wave packet is gaussian.
}
\end{figure}

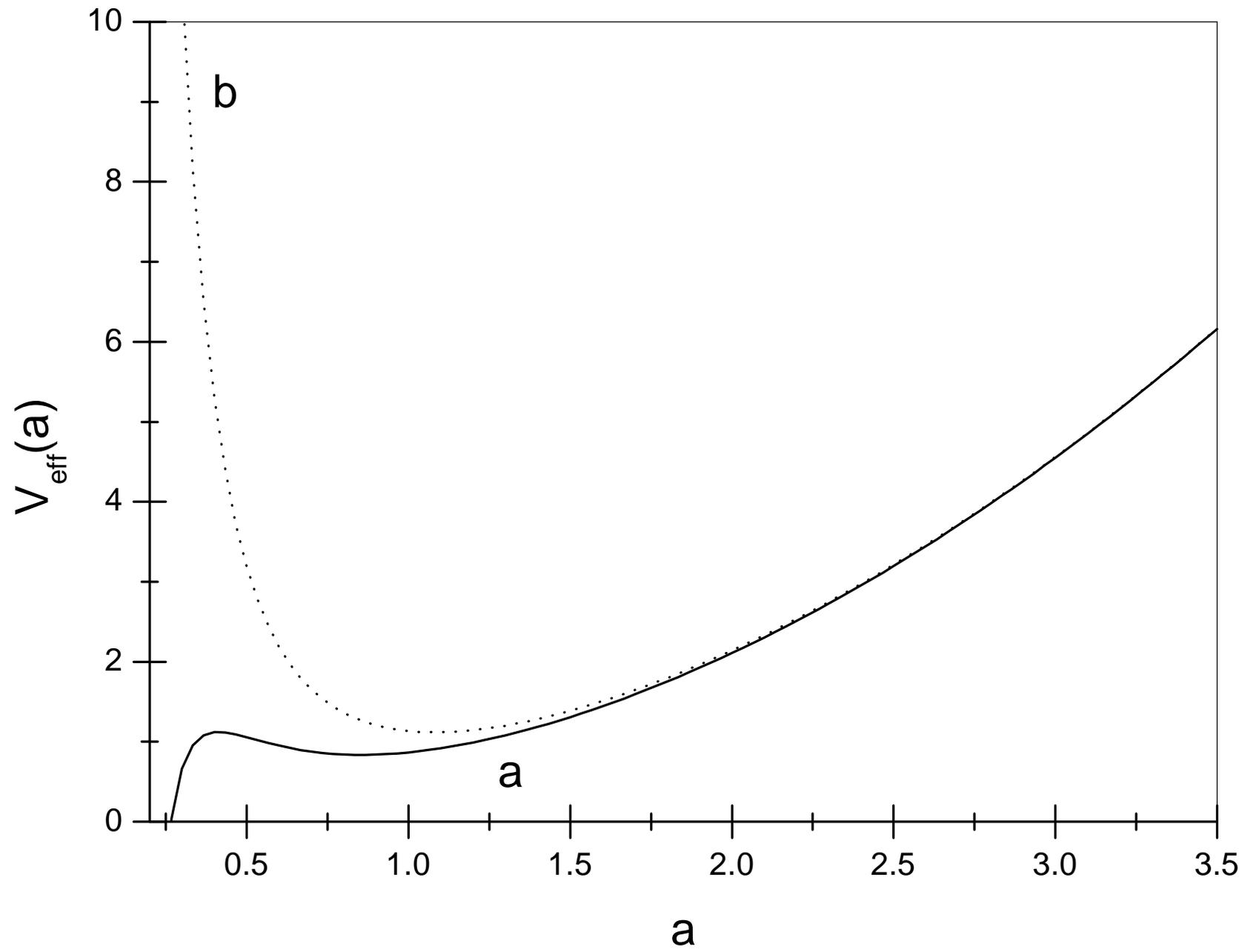
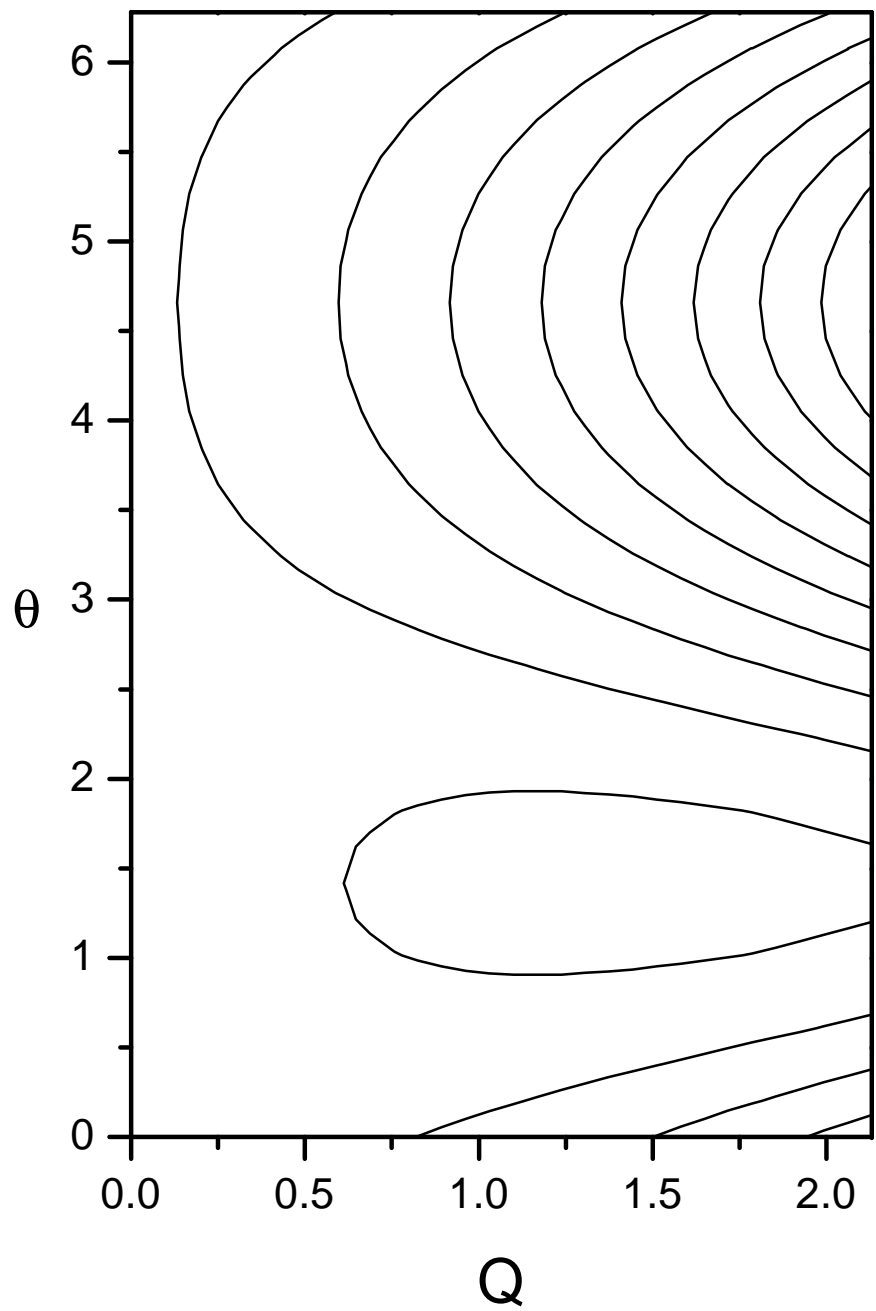
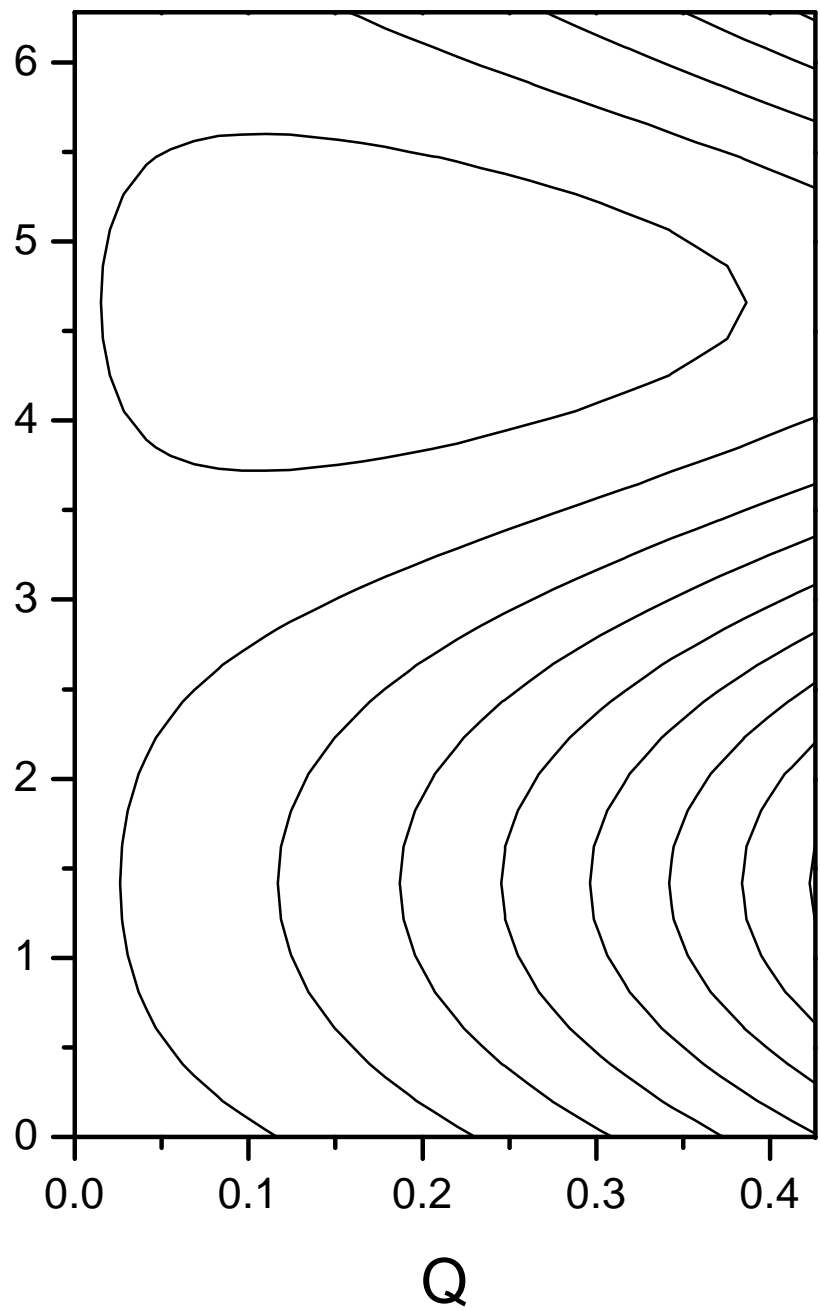
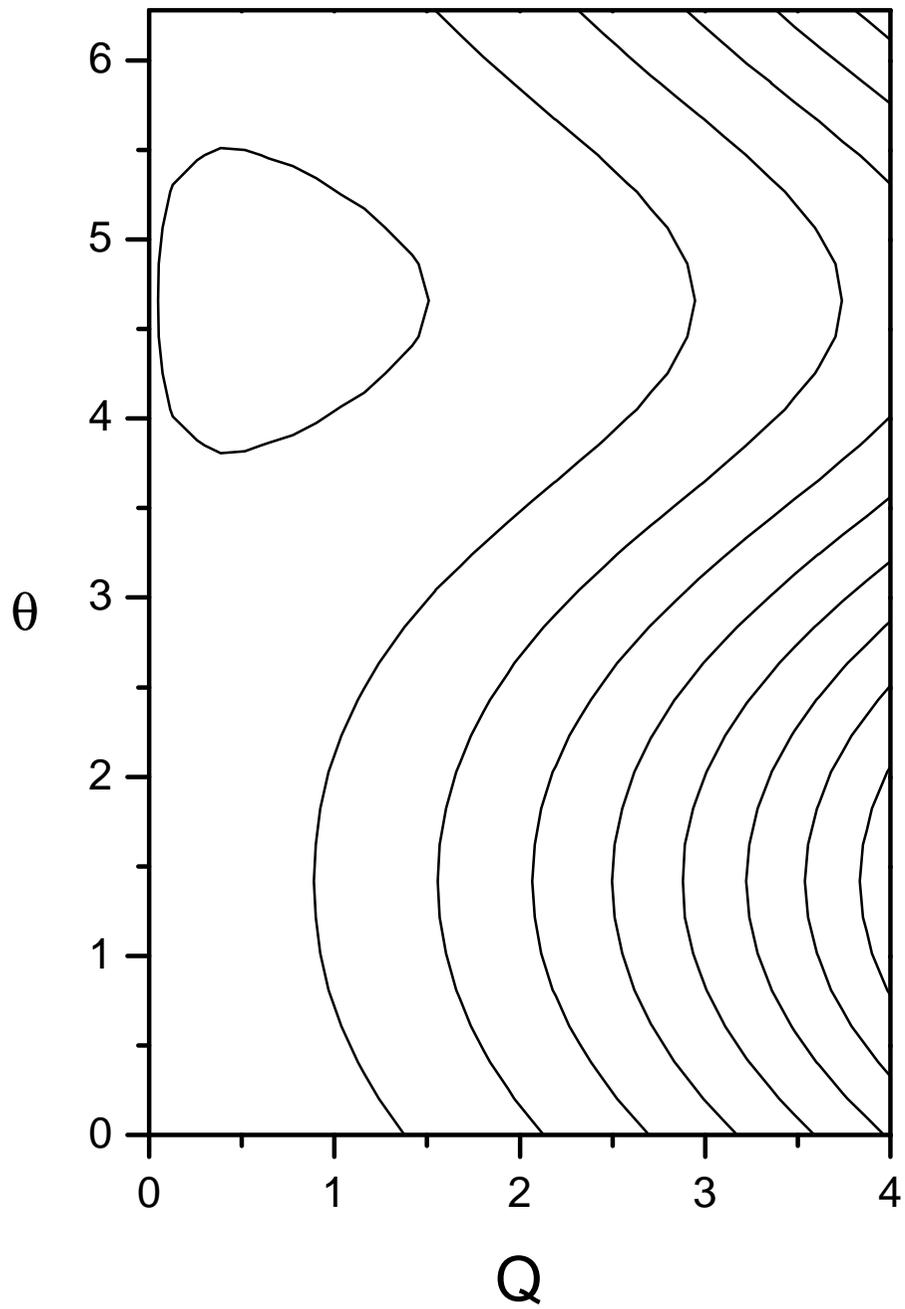
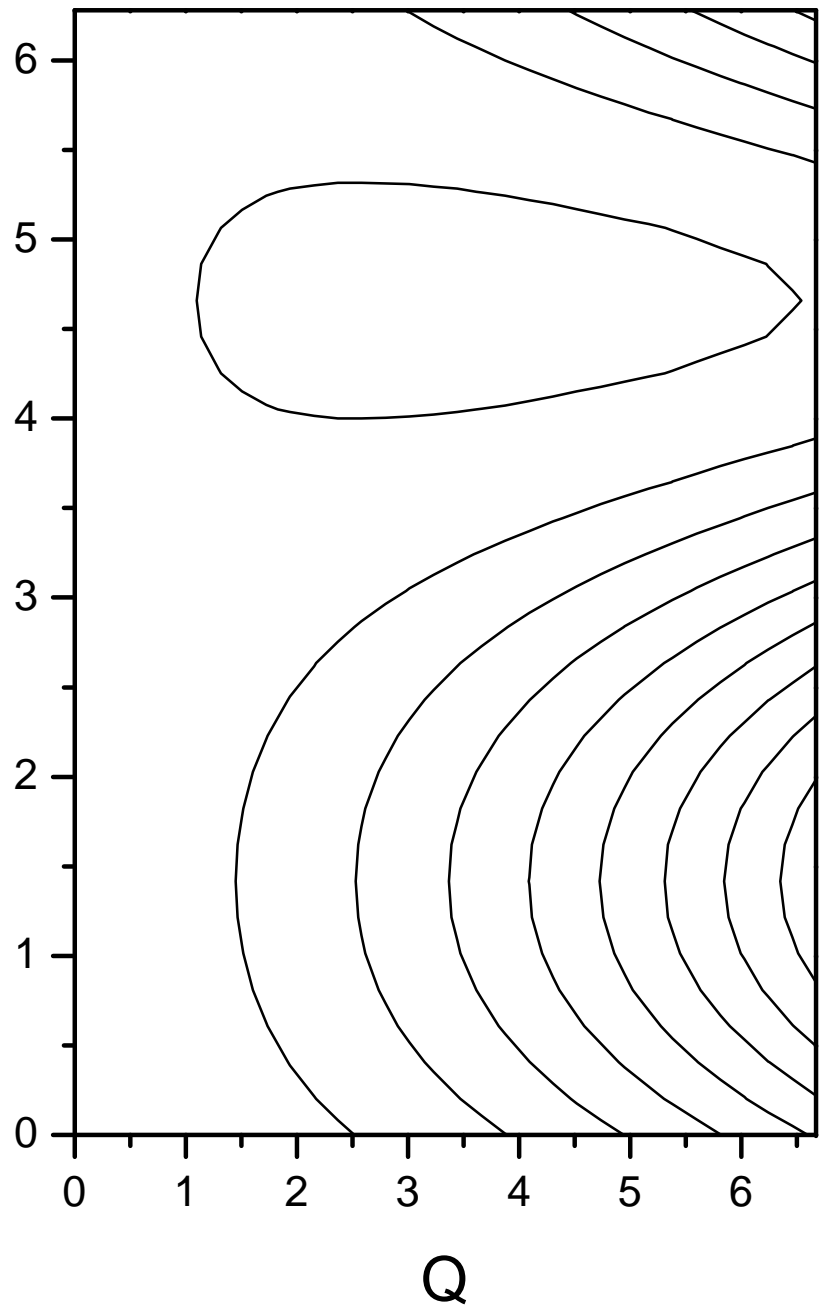
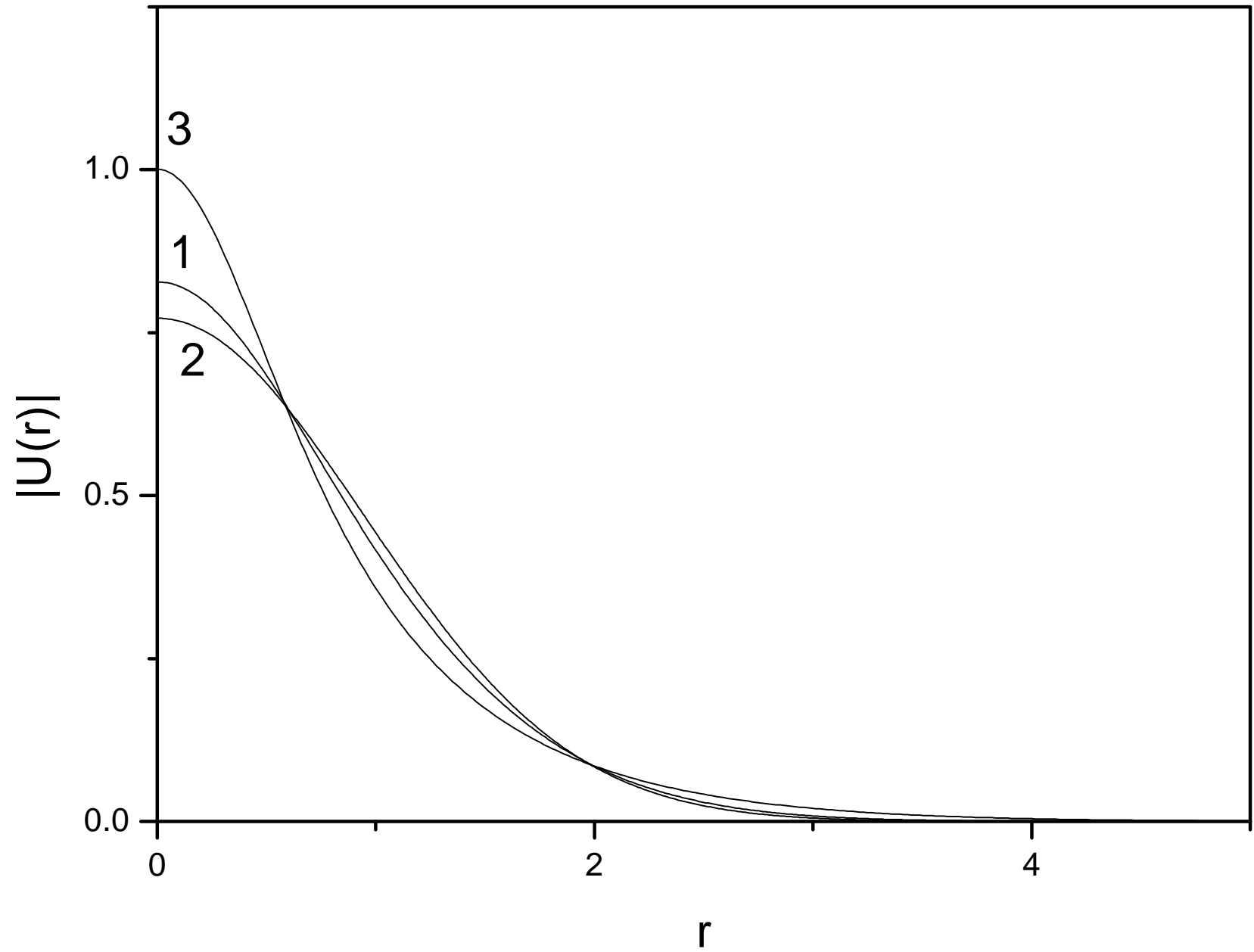
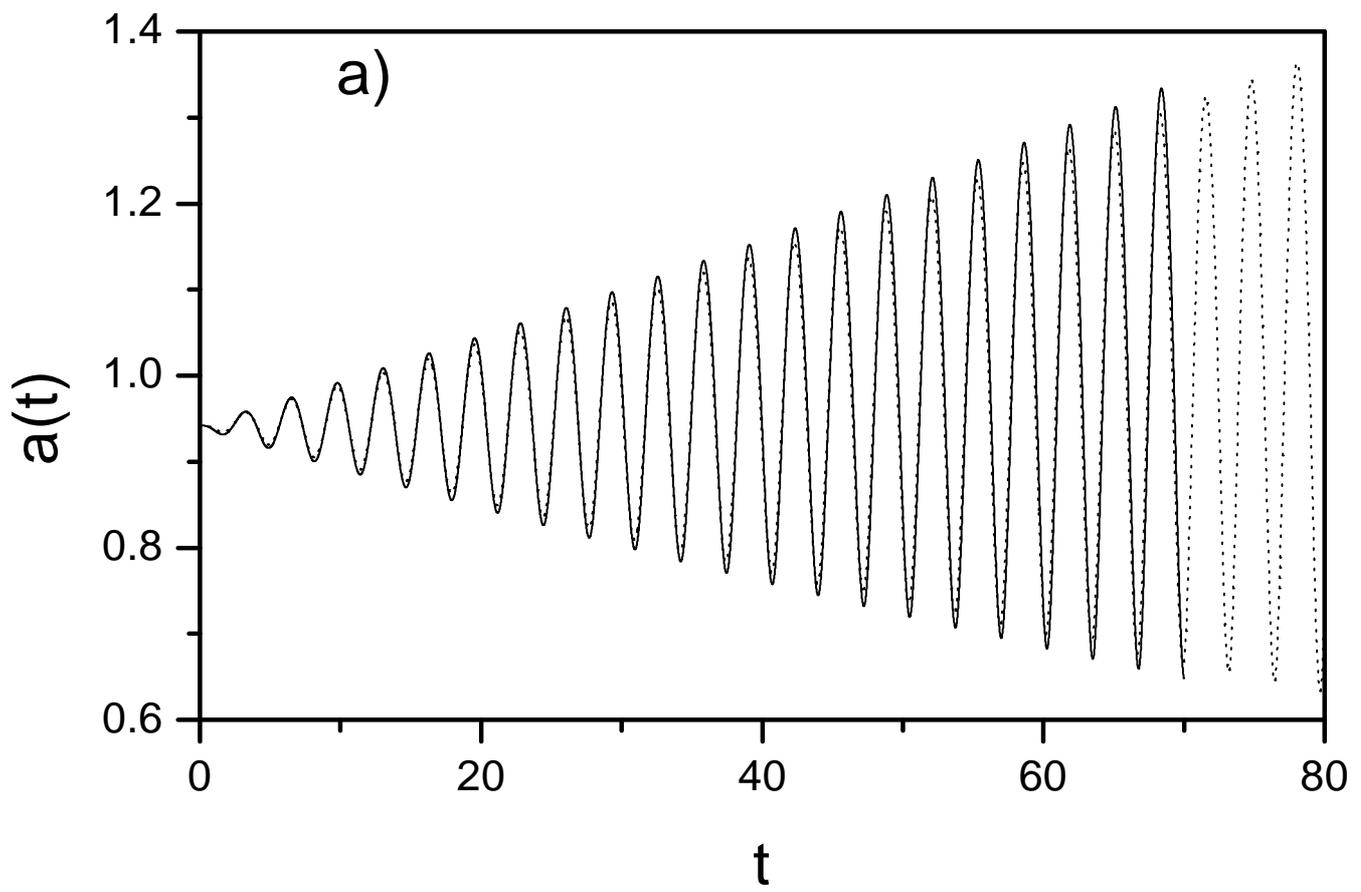
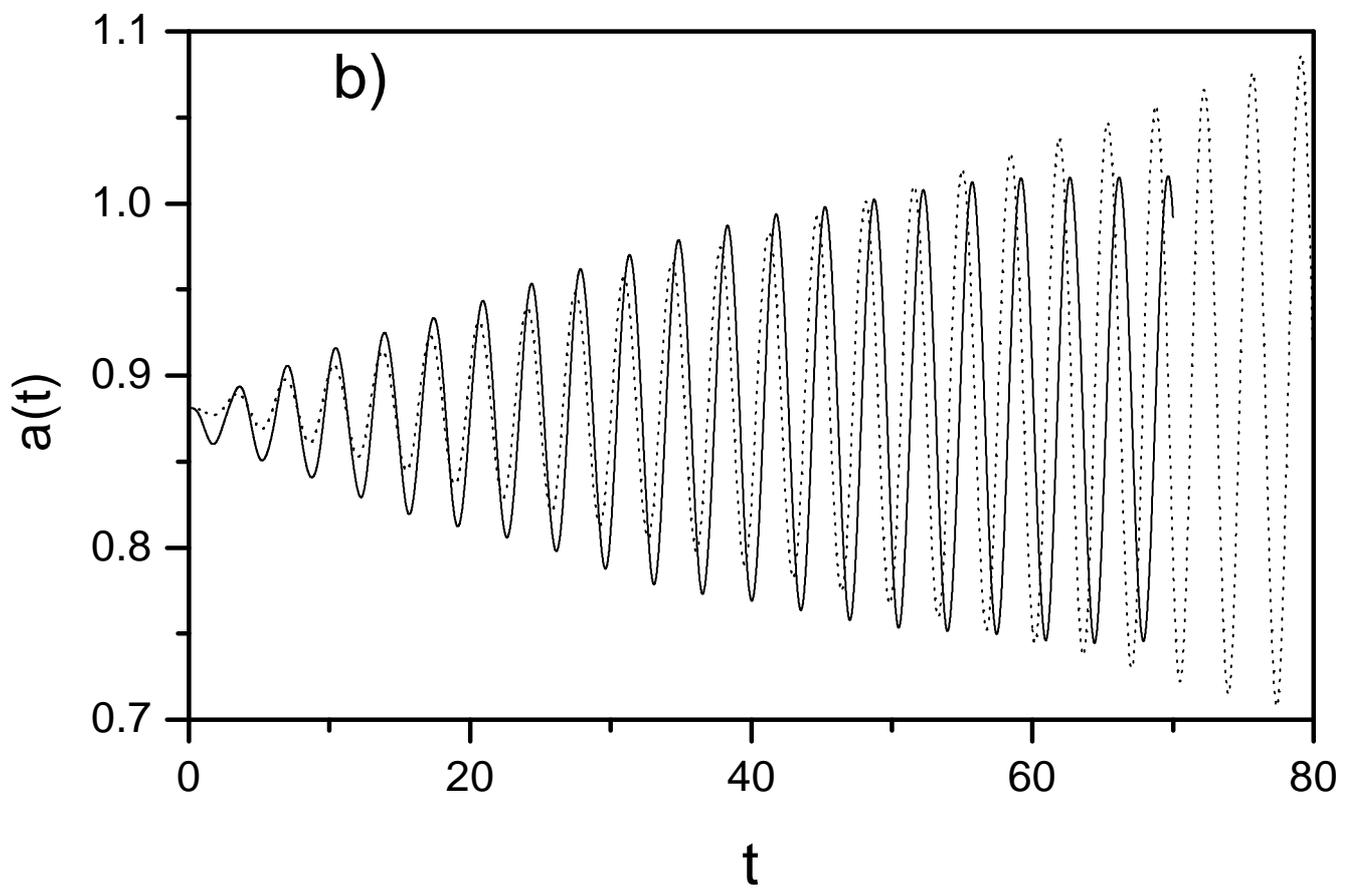
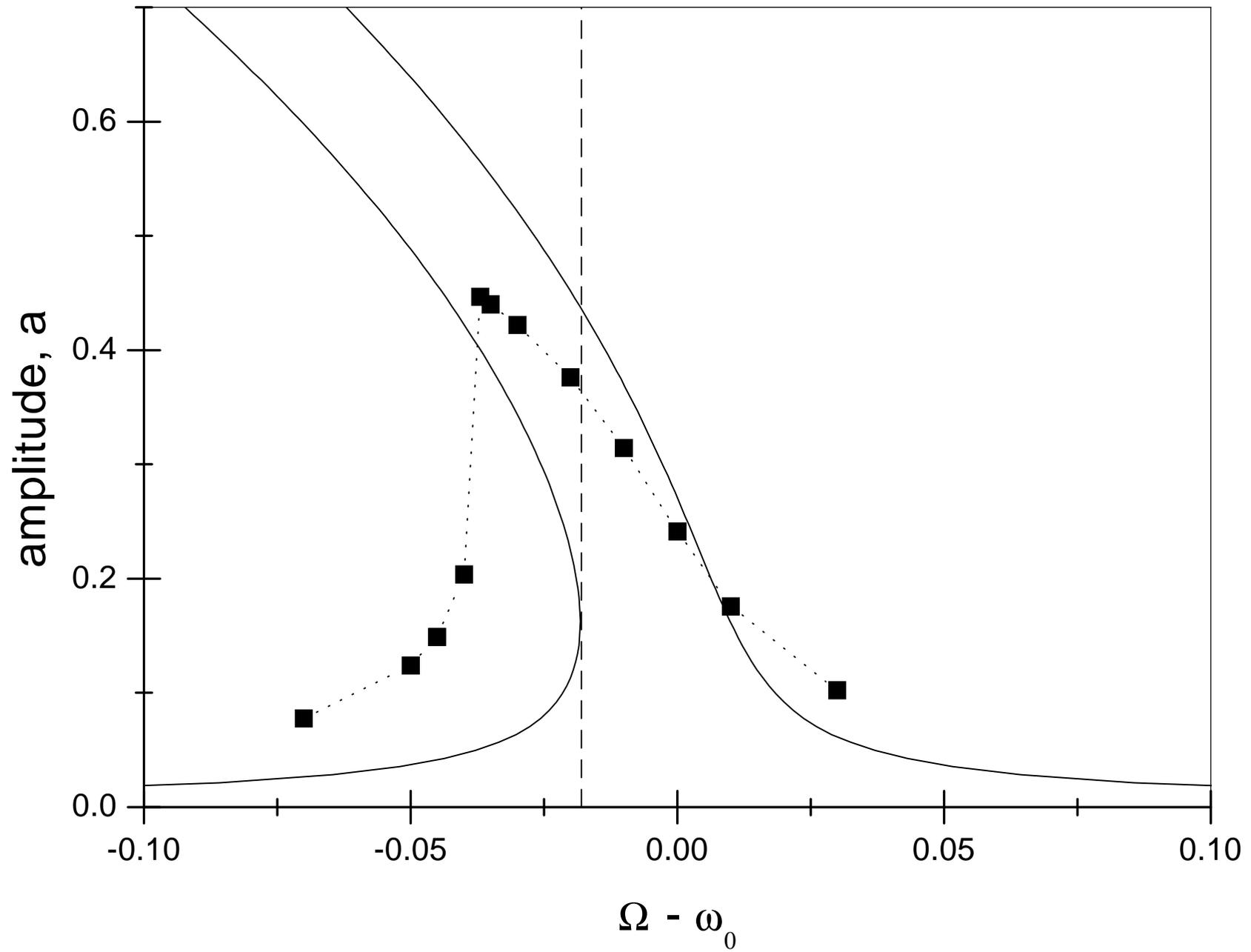
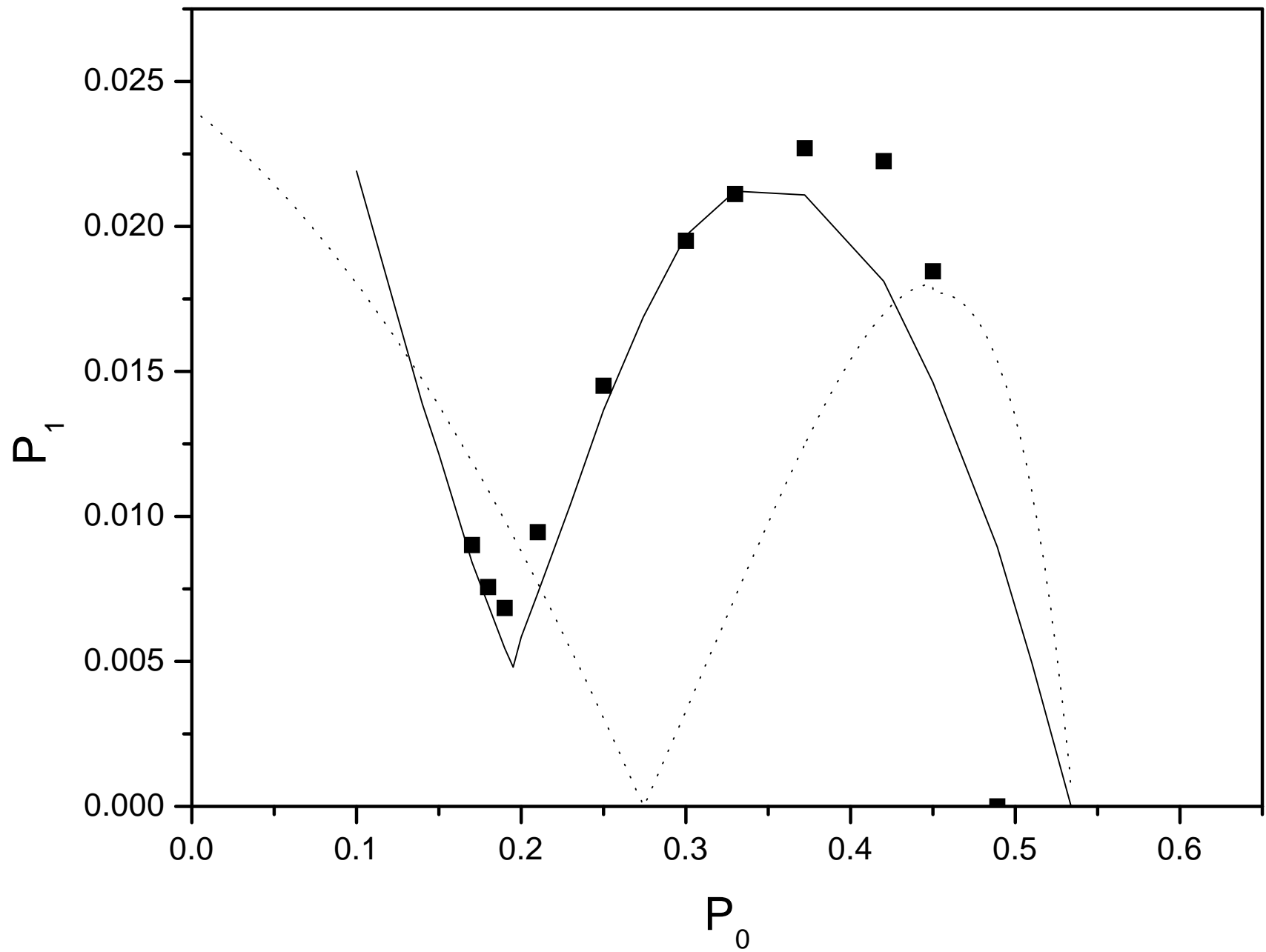
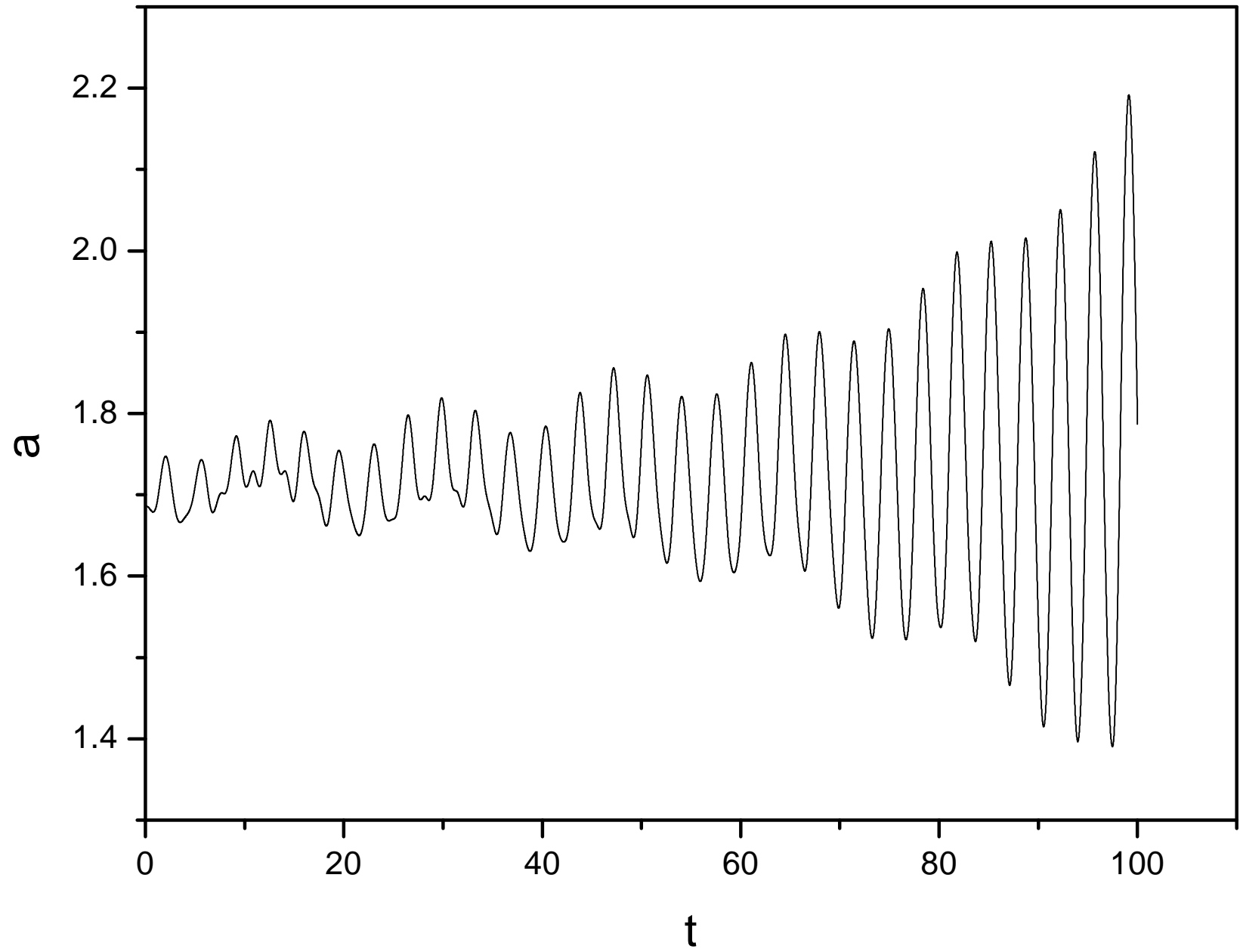
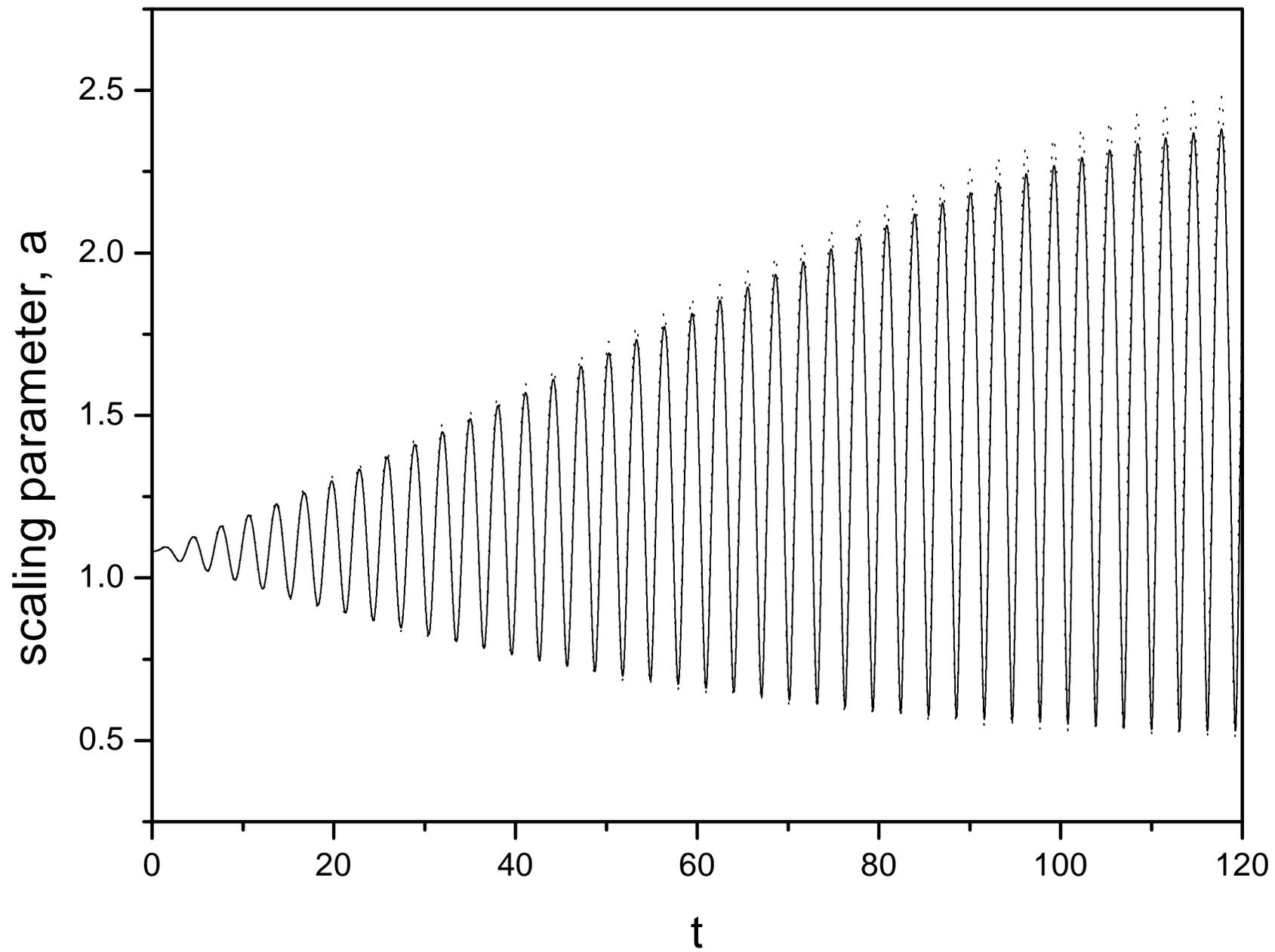
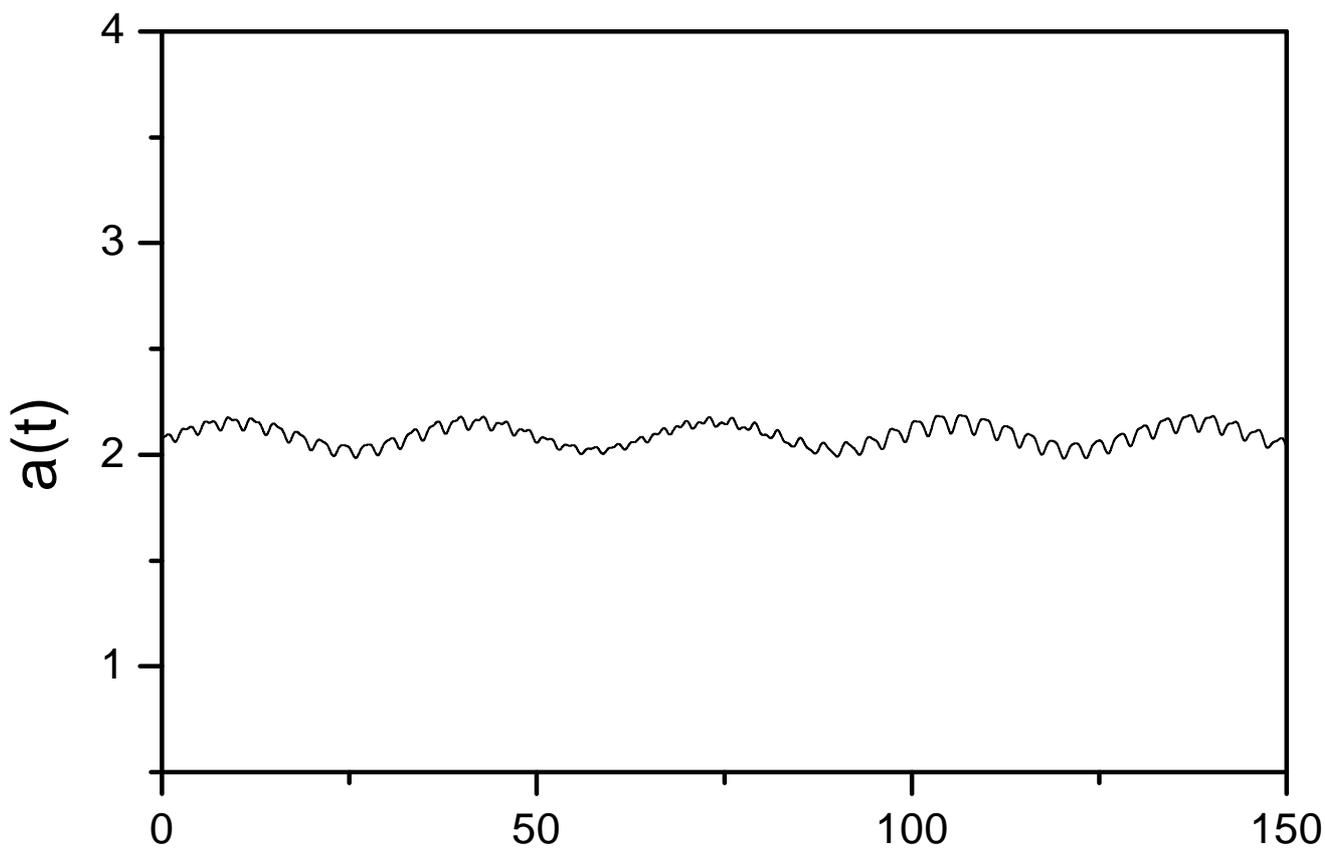
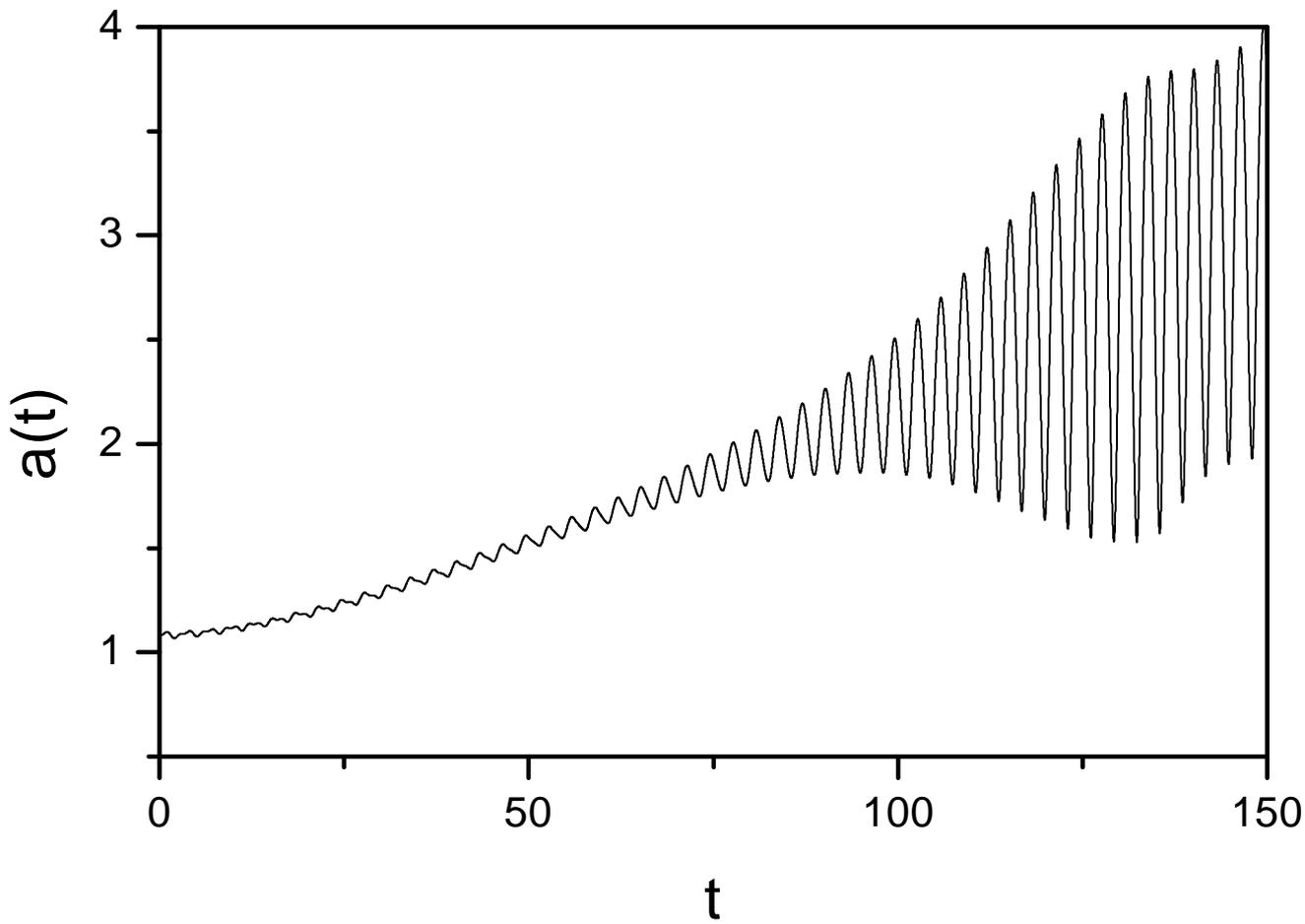
\end{document}